\pdfoutput=1 
\documentclass[%aip,apl,reprint
 floatfix,
superscriptaddress,
 showkeys,
 aps,
reprint
]{revtex4-1}
\usepackage{mathtools}
\usepackage{array}
\usepackage{tabularx}
\usepackage{graphicx}
\usepackage{dcolumn}% Align table columns on decimal point
\usepackage{bm}% bold math
\usepackage{amsmath}%[fleqn]
\usepackage{color}
\usepackage{wasysym}
\usepackage{float}
\usepackage{array,url,kantlipsum}
\usepackage{verbatim}
\usepackage{xcolor}
\usepackage{CJK}
\usepackage{textcomp}
\newcolumntype{P}[1]{>{\centering\arraybackslash}p{#1}}
\usepackage{array}
\usepackage{amssymb}
\newcolumntype{x}[1]{>{\centering\arraybackslash\hspace{0pt}}p{#1}}

\begin{document}
\title{Fiberized diamond-based vector magnetometers}
\author{Georgios Chatzidrosos}
\affiliation{Helmholtz-Institut, GSI Helmholtzzentrum f{\"u}r Schwerionenforschung, 55128 Mainz, Germany}
\affiliation{Johannes Gutenberg-Universit{\"a}t Mainz, 55128 Mainz, Germany}
\email[corresponding authors: \\]{gechatzi@uni-mainz.de, zheng@uni-mainz.de}
\author{Joseph Shaji Rebeirro}
\affiliation{Helmholtz-Institut, GSI Helmholtzzentrum f{\"u}r Schwerionenforschung, 55128 Mainz, Germany}
\affiliation{Johannes Gutenberg-Universit{\"a}t Mainz, 55128 Mainz, Germany}
\author{Huijie Zheng}
\affiliation{Helmholtz-Institut, GSI Helmholtzzentrum f{\"u}r Schwerionenforschung, 55128 Mainz, Germany}
\affiliation{Johannes Gutenberg-Universit{\"a}t Mainz, 55128 Mainz, Germany}
\author{Muhib Omar}
\affiliation{Helmholtz-Institut, GSI Helmholtzzentrum f{\"u}r Schwerionenforschung, 55128 Mainz, Germany}
\affiliation{Johannes Gutenberg-Universit{\"a}t Mainz, 55128 Mainz, Germany}
\author{Andreas Brenneis}
\affiliation{Corporate Sector Research and Advance Engineering, Robert Bosch GmbH, D-71272 Renningen, Germany}
\author{Felix M. Stürner}
\affiliation{Corporate Sector Research and Advance Engineering, Robert Bosch GmbH, D-71272 Renningen, Germany}
\author{Tino Fuchs}
\affiliation{Corporate Sector Research and Advance Engineering, Robert Bosch GmbH, D-71272 Renningen, Germany}
\author{Thomas Buck}
\affiliation{Corporate Sector Research and Advance Engineering, Robert Bosch GmbH, D-71272 Renningen, Germany}
\author{Robert Rölver}
\affiliation{Corporate Sector Research and Advance Engineering, Robert Bosch GmbH, D-71272 Renningen, Germany}
\author{Tim Schneemann}
\affiliation{Helmholtz-Institut, GSI Helmholtzzentrum f{\"u}r Schwerionenforschung, 55128 Mainz, Germany}
\affiliation{Johannes Gutenberg-Universit{\"a}t Mainz, 55128 Mainz, Germany}
\author{Peter Blümler}
\affiliation{Johannes Gutenberg-Universit{\"a}t Mainz, 55128 Mainz, Germany}
\author{Dmitry Budker}
\affiliation{Helmholtz-Institut, GSI Helmholtzzentrum f{\"u}r Schwerionenforschung, 55128 Mainz, Germany}
\affiliation{Johannes Gutenberg-Universit{\"a}t  Mainz, 55128 Mainz, Germany}
\affiliation{Department of Physics, University of California, Berkeley, CA 94720-7300, USA}
\author{Arne Wickenbrock}
\affiliation{Helmholtz-Institut, GSI Helmholtzzentrum f{\"u}r Schwerionenforschung, 55128 Mainz, Germany}
\affiliation{Johannes Gutenberg-Universit{\"a}t  Mainz, 55128 Mainz, Germany}
\date{\today}
\begin{abstract}
We present two fiberized vector magnetic-field sensors, based on nitrogen-vacancy (NV) centers in diamond. The sensors feature sub-nT/$\sqrt{\textrm{Hz}}$ magnetic sensitivity. We use commercially available components to construct sensors with a small sensor size, high photon collection, and minimal sensor-sample distance. Both sensors are located at the end of optical fibres with the sensor-head freely accessible and robust under movement.  These features make them ideal for mapping magnetic fields with high sensitivity and spatial resolution ($\leq$\,mm).  As a demonstration we use one of the sensors to map the vector magnetic field inside the bore of a $\geq$ 100\,mT Halbach array. The vector field sensing protocol translates microwave spectroscopy data addressing all diamonds axes and including double quantum transitions to a 3D magnetic field vector.
\end{abstract}
\maketitle

\section*{Introduction}

\begin{figure*}
    \centering
    \includegraphics[width=1\textwidth]{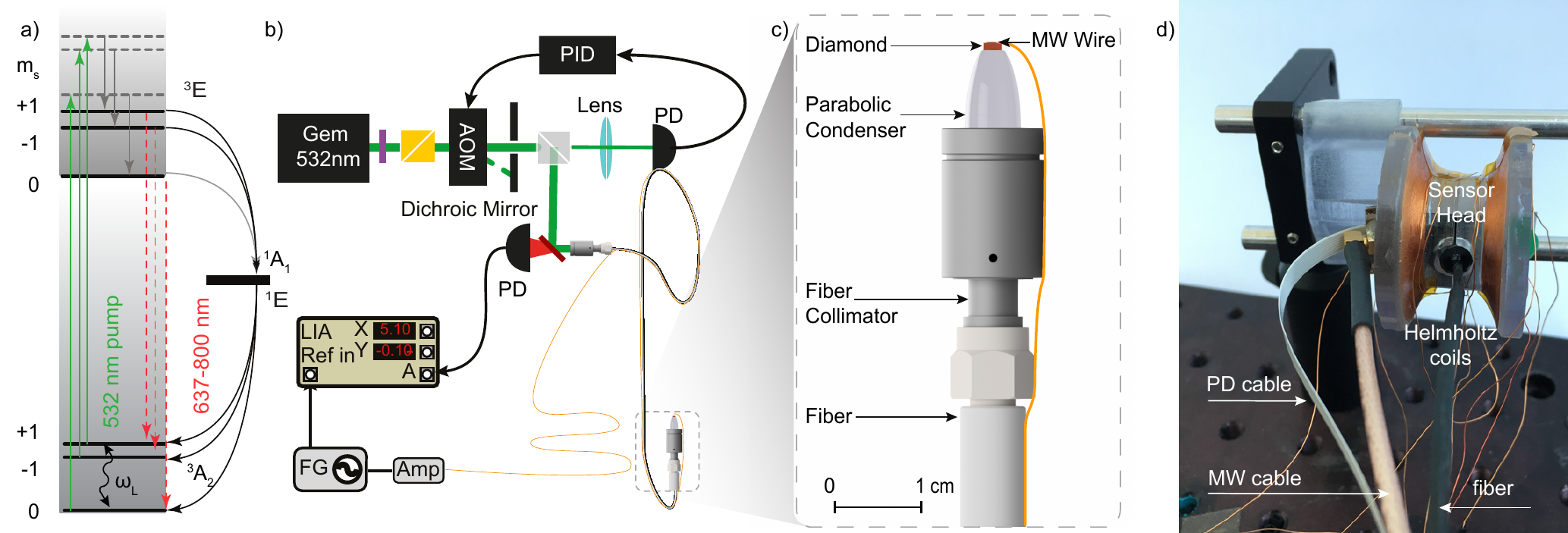}
    \caption{a) Relevant energy levels of the NV center. The states as well as their spin projection m$_s$ are labeled. Solid green lines represent pump light, solid gray lines phononic de-excitation, dashed red lines represent PL, curved gray lines represent inter-system crossing. b) Experimental setup in Mainz (the experimental setup by Bosch can be found in Ref.\,\cite{Boscharxiv}). c) Schematic of the fiberized sensor head. d) Photograph of the fiberized sensor by Robert Bosch GmbH. The abbreviations used in the figure are; PD: Photo-diode, AOM: acousto-optic modulator, PID: proportional-integral-derivative controller, LIA: lock-in amplifier, FG:  function generator, Amp: amplifier, MW: microwave.}
    \label{fig:1}
\end{figure*}

Nitrogen-vacancy (NV) centers in diamond have attracted attention as magnetic field sensors with high spatial resolution\,\cite{Balasubramanian2008,Maze2008,Rittweger2009} and sensitivity\,\cite{Barry2016,Chatzidrosos}.
% Prominent examples. 
The range of their application includes, but is not limited to, single neuron-action potential detection\,\cite{Barry2016}, single protein spectroscopy\,\cite{Lovchinsky2016}, as well as in-vivo thermometry\,\cite{SingleNVThermometry}. Advantages of NV-based magnetometers, compared to other magnetic field sensors, include their ability to operate in wide temperature and magnetic field ranges\,\cite{levelanticrossing}. The ability to operate them also without the use of microwaves\,\cite{wickenbrock2016microwave,levelanticrossing,Zheng2020}, has recently enabled a variety of new applications in environments where microwaves (MW) would be detrimental\,\cite{Chatzi2019eddycurrent}.

The NV consists of a substitutional nitrogen and an adjacent vacant carbon site. It can appear in different orientations along the crystallographic axes of the diamond lattice. This enables vector measurements of magnetic fields\,\cite{vectorClevenson}. Vector magnetometry itself can be useful in magnetic navigation applications\,\cite{Cochrane}, magnetic
anomaly detection, current and position sensing, and the measurement of biological magnetic fields\,\cite{Barry2016}. Vector measurements near a background field of $\sim$\,100\,mT, where the NV's ground state level anti-crossing (GSLAC) occurs, are of particular interest\,\cite{Zheng2020}. Some of the challenges of vector measurements near the GSLAC field include the necessity to precisely align the NV and the bias magnetic field axes for optimum sensitivity of the microwave-free method \cite{levelanticrossing} or the need to account for transversal field related nonlinearities of the NV gyromagnetic ratio when performing microwave spectroscopy. 

One of the challenges NV magnetometers face is their low photon-collection efficiency.  Approaches to increase the efficiency include, use of solid immersion lenses \cite{Hadden2010,Siyushev2010,Sage2012}, or employment of infrared absorption\,\cite{Dumeige2013,Budker1,Jensen2014, Chatzidrosos, intraetalon}. Photoluminescence (PL) for fiberized sensors is preferentially collected with the same fiber delivering the pump light but detected on the input side of the fiber\,\cite{fibersenso}. Despite considerable effort, even modern sensors typically just feature a PL-to-pump-light ratio of about 0.1\%\,\cite{fibersenso,Barry2016}. 

In this paper, we discuss the construction of two fiberized NV-based vector magnetic field sensors one constructed in Helmholtz Institute Mainz, referred to as Mainz sensor in the following and the other one is a sensor demonstrator of Robert Bosch GmbH, referred to as Bosch sensor in the following. They both achieve sub-nT$/\sqrt{\textrm{Hz}}$ magnetic sensitivity. The sensor constructed in Mainz achieved a sensitivity of 453\,pT/$\sqrt{\textrm{Hz}}$, limited by intensity noise of the pump laser, with 11\,pT/$\sqrt{\textrm{Hz}}$ photon-shot noise sensitivity and 0.5\% PL-to-pump power ratio taking the fiber coupling efficiency into account. The sensor constructed by Bosch achieved a sensitivity of 344\,pT/$\sqrt{\textrm{Hz}}$, which was approximately one order of magnitude larger than the expected photon-shot noise limited sensitivity\,\cite{Boscharxiv}. The PL-to-pump power ratio was 0.3\% taking the fiber-coupling efficiency into account. The components used for the construction of both sensors are commercially available. They allow for a small sensor size (11\,$\times\,$40\,mm$^2$ for Mainz and 15\,$\times$\,25\,mm$^2$ for Bosch) with maximized PL-to-pump-light ratio, as well as robustness to movement, which also makes the sensors portable. All of this makes the sensors ideal for mapping magnetic fields and measuring in regions that are not easily accessible. The magnetometers are constructed in such a way that they allow close proximity 
%($\leq$\,300\,$\mu$m \textcolor{blue}{(If this is not true now we should remove this sentence)}) 
of the sensors to magnetic field sources, thus allowing for high spatial resolution when mapping magnetic fields.
After demonstrating the sensitivity of the sensors and explaining the principles of vector magnetometry with NV centers,  we used the Mainz sensor to
%as a demonstration of the versatility of this sensors,
perform spatially resolved optically detected magnetic resonance (ODMR) measurements covering an
area of $20\times30\,\text{mm}^2$ inside a Halbach-magnet array. The Halbach array itself provides a highly homogeneous magnetic field around 100\,mT which makes it ideal for near-GSLAC magnetic field measurements and studies with NV centers. We present the analysis for the translation of the extracted frequency measurements into magnetic field. Details about the construction of the highly homogeneous Halbach array is subject of another publication\,\cite{Halbach}.

\section*{Experimental Setup}
% Diamond description

The diamond sample used for the Mainz sensor is a 2.0\,$\times$\,2.0\,$\times$\,0.5\,mm$^3$ type Ib, (100)-cut, high-pressure high-temperature (HPHT) grown sample, purchased from Element Six. The initial [N] concentration of the sample was specified as $<$\,10\,ppm. The sample was irradiated with 5\,MeV electrons at a dose of $2\times 10^{19}\,\textrm{cm}^{-2}$ and then annealed at 700\,$^{\circ}$C for 8\,hours.
The diamond sample used for the Bosch sensor is a $0.8\,\times\,0.8\,\times\,0.5\, \textrm{mm}^3$ (111)-cut, 99.97\,\% $^{12}$C enriched, HPHT grown diamond. The sample was irradiated with 2\,MeV electrons at a dose of $2\,\times\,10^{18}\,\textrm{cm}^{-2}$ at room temperature and then annealed at 1000\,$^{\circ}$C for 2\,hours in vacuum. The [NV$^-$] concentration was determined by electron spin resonance to be 0.4\,ppm\,\cite{Boscharxiv}.

% Fig1(a) intro. Energy level structure.
% Present the energy levels. triplet.
Figure\,\ref{fig:1}\,(a) shows the relevant energy levels of the negatively charged NV center, which we use for magnetometry.
The ground and excited spin-triplet states of the NV are labeled $^{3}$A$_{2}$ and $^{3}$E, respectively [Fig.\,\ref{fig:1}\,(a)], the transition between them has a zero-phonon line at 637\,nm, but can be exited by more energetic photons (of shorter wavelengths) due to phonon excitation in the diamond lattice. 
% Present the energy levels. singlet.
The lower and upper electronic singlet states are $^{1}$E and $^{1}$A$_{1}$, respectively.
%with the transition between them having a zero-phonon line at 1042\,nm. 
% Explaining the picture: lines. Spin dependence.
While optical transition rates are spin-independent, the probability of nonradiative intersystem crossing from $^3$E to the singlets is several times higher\,\cite{Dumeige2013} for $m_s=\pm1$ than that for $m_s=0$. 
% Optical pumping. Step one Prepare
As a consequence, under continuous illumination with green pump light (532\,nm), NV centers accumulate mostly in the $^{3}$A$_{2}$, $m_s=0$ ground state sublevel and in the metastable $^{1}$E singlet state. 
% MW. Step two Drive.
For metrology applications, the spins in the $^{3}$A$_{2}$ ground state can be coherently manipulated with microwave fields.

% What we do to Green light
Figure\,\ref{fig:1}\,(b) shows the experimental setup used for the measurements conducted in Mainz.
To initialize the NV centers to the ground-triplet state, we use a 532\,nm (green) laser (Laser Quantum, gem532). To reduce intensity noise from the laser source, the light intensity is stabilized using an acousto-optic modulator (AOM, ISOMET-1260C with an ISOMET 630C-350 driver) controlled through a proportional-integral-derivative controller (PID, SIM960), in a feedback loop. The green laser is then coupled into a 2\,m long, high-power multi-mode fiber cable (Thorlabs, MHP365L02).

The MW to manipulate the NV spins are generated with a MW function generator (FG; SRS, SG394). They are amplified with a 16\,W amplifier (Amp; Mini-Circuits, ZHL-16W-43+) and passed through a circulator (Mini-Circuits, CS-3.000) before they are applied to the NV centers using a mm-sized wire loop. The other side of the wire is directly connected to ground. The radius of the wire used for the wire loop is 50\,$\mu$m. The wire is then attached to the optical fiber allowing the two parts to move together. For the magnetic resonance measurements presented in this paper, the MW frequency was scanned between 3800\,MHz and 5000\,MHz. Information on the experimental setup used to conduct the measurements with the Bosch sensor can be found in Ref\,\cite{Boscharxiv}.

\begin{table}
  \centering
\begin{tabular}{|m{1.8cm}|m{1.3cm}|m{1.5cm}|m{2.8cm}|}
 \hline
$f_{res}$ &$f_{mod}$ &$f_{depth}$& Detection limit\\
 \hline
 \hline
2880.4\,MHz & 3\,kHz & 100\,kHz & 3.7\,$\pm$\,0.6\,nT/$\sqrt{\textrm{Hz}}$ \\
 \hline
2894.3\,MHz & 2\,kHz &  100\,kHz & 3.1\,$\pm$\,0.3\,nT/$\sqrt{\textrm{Hz}}$\\
\hline
2873.5\,MHz  & 5\,kHz  & 100\,kHz  & 2.7\,$\pm$\,1.6\,nT/$\sqrt{\textrm{Hz}}$ \\\hline
\end{tabular}
\caption{Settings and detection limit for the three different NV axis used in the Bosch sensor.}
\label{table1}
\end{table}

A schematic of the Mainz fiberized magnetic field sensor head is shown in Fig.\,\ref{fig:1}\,(c). The diamond is glued to a  parabolic condenser lens, which itself is glued onto a 11\,mm focal length lens and attached to a fiber collimator (Thorlabs, F220SMA-532) to which the high-power multi-mode fiber (Thorlabs, MHP365L02) is connected. The MW wire loop is attached to the other side of the diamond to provide the rapidly oscillating magnetic fields required for this magnetic field detection scheme. The light is delivered to the diamond via the parabolic condenser, lens, collimator and the high-power fiber. The same components collect the spin-state-dependent red PL of the NV ensemble. On the other side of the fibre the PL is filtered using a longpass dichroic filter (Thorlabs, DMLP605) which is also used to couple the incoming green light into the fiber. After the dichroic filter residual reflected green light is removed by a notch filter (Thorlabs, NF533-17) and the PL is focused onto a photodiode (PD; Thorlabs, PDA36a2). The detected signal with the PD is connected to a a lock-in amplifier (LIA; SRS, SR830). With this setup we were able to achieve 0.5\% PL-to-pump-light ratio, which is an order-of-magnitude improvement compared to other fiberized sensors \cite{fibersenso}. 
A photograph of the Bosch sensor can be seen in Fig.\,\ref{fig:1}\,(d). The sensor head, containing a microwave resonator, a custom designed balanced photodetector, and the diamond, which was glued to the collimated output of a single-mode fiber, is located inside a custom designed Helmholtz coil. The Helmholtz coil was used to generate a magnetic bias field of 1.07\,mT and the collimation of the laser beam was achieved with a gradient-index lens (GRIN). Further details on the setup used for the Bosch sensor can be found in Ref.\,\cite{Boscharxiv}.
The final sensor head, of the Mainz sensor, has a diameter of $\leq$\,11\,mm and a height of 40\,mm, in a configuration that allows for $\leq$\,300\,$\mu$m average distance between sample and sensor. The 2\,m long fiber with the attached MW wire and the fiberized sensor head for the Mainz sensor (shown in Fig.\,\ref{fig:1}\,(c)) can be moved independently of the other components. The footprint of the sensor head of the Bosch sensor was 15\,$\times$25\,mm$^2$. The smallest distance between the center of the diamond and the outer surface is roughly 2.4\,mm, the sensor can move independently as well. %The components used for the two sensors are commercially available.
To filter low-frequency systematic noise components, e.g.  laser-power fluctuations, the MW-frequency $\omega_L$ is modulated and the detected PL is demodulated with the LIA. 
To produce the magnetic field maps presented here, the Mainz sensor assembly was mounted on a computer-controlled motorized 3D translation stage (Thorlabs, MTS25/M-Z8) in the center of the Halbach array (not shown in Fig.\,\ref{fig:1}).

\section*{Magnetic field sensitivity}

%\begin{figure}
%    \centering
%    \includegraphics[width=1\columnwidth]{Figures/ODMRfiberize2.pdf}
%    \caption{ODMR of NVs aligned along a single axis. Three peaks are visible due to hyperfine interactions with the $^{14}$N nuclear spin. The average width of the features is 350(3)\,kHz, and the contrast is 1.6(1)\% }
%   \label{fig:2}
%\end{figure}

\begin{figure}
    \centering
    \includegraphics[width=1\columnwidth]{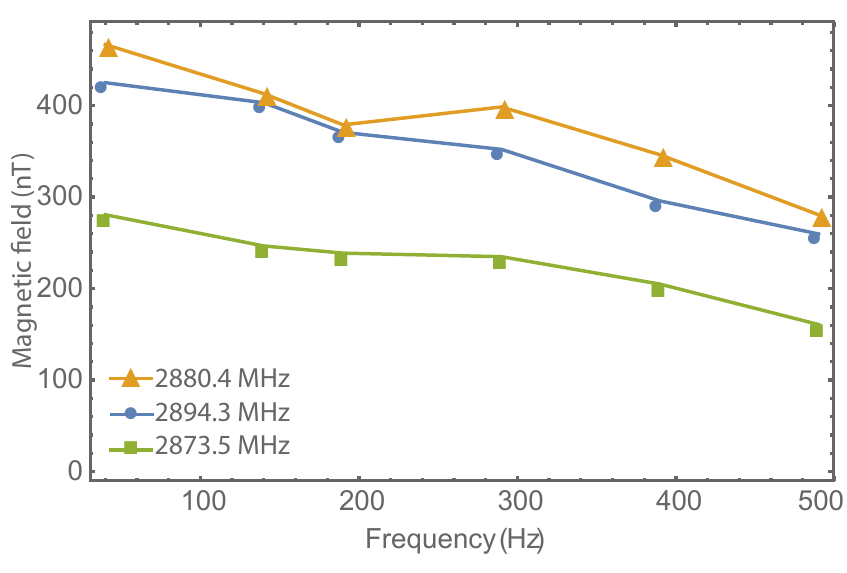}
    \caption{Amplitude of magnetic test signal, measured along three different NV axes with the Bosch sensor. An oscillating, external magnetic field of about 0.7 µT was applied to the sensor.}
   \label{fig:2}
\end{figure}

% peaks that appear
Figure\,\ref{fig:2} depicts the response of the Bosch sensor for three different NV axes upon application of an AC magnetic field with varying frequencies and a constant amplitude of 0.7\,$\mu$T. The three axes see different effective amplitudes due to the different angles of the NV axes with respect to the magnetic field vector as expected. The sensor response slightly decreases with increasing frequency of the test field independent of the axis. This decrease might stem from the fact, that with increasing frequency, the excitation field becomes under-sampled meaning that the full reconstruction of the sinusoidal excitation field is not possible anymore leading to a decrease in measured amplitude.

To estimate the magnetic sensitivity of the two sensors we follow another method. When the MW field is resonant with the ground-state $m_s=0$ $\rightarrow$ $m_s=\pm 1$ transitions, population is transferred through the excited triplet state to the metastable singlet state, resulting in PL reduction. PL as a function of MW frequency is the ODMR signal. 
%Figure\,\ref{fig:2} shows the ODMR for a single NV orientation used for this experiment under an external magnetic field of $\sim$\,0.5\,mT. Three peaks are visible due to hyperfine interactions with the $^{14}$N nuclear spins, this is omitted from Fig\,\ref{fig:1}\,(a). 
%The average width of the features is 350\,(3)\,kHz, and the contrast 1.6\,(1)\%. 
%To estimate the magnetic field sensitivity of the sensor, we scan the MW field around
%one of the features displayed in Fig.\,\ref{fig:2}. 
%a single ODMR feature signal.
By focusing around a single of feature in this ODMR signal the sensitivity can be extracted.
The magnetic resonance for the Mainz sensor features a 350\,kHz linewidth and 1.6\,\% contrast. The best sensitivity of the Bosch sensor was achieved with a linewidth of $\approx$ 92 kHz and a contrast of 0.6 \%.
We modulate the MW frequency around a central frequency f$_{c}$, and record the first harmonic of the transmission signal with a LIA. This generates a demodulated signal, which together with the NV gyromagnetic ratio of $|^{\gamma}/_{2\pi}| \sim 28.024\,\rm{GHz}\,\rm{T}^{-1}$  can be used to translate PL fluctuations to effective magnetic field noise. The optimized parameters resulting in the best magnetic field sensitivity for the Mainz sensor were: a modulation frequency of f$_{mod}$ = 13.6\,kHz and modulation amplitude of f$_{amp}$ = 260\,kHz. The sensitivity of the Bosch sensor was optimized for 3 different NV axes as summarized in table\,\ref{table1}.

\begin{figure}
    \centering
    \includegraphics[width=1\columnwidth]{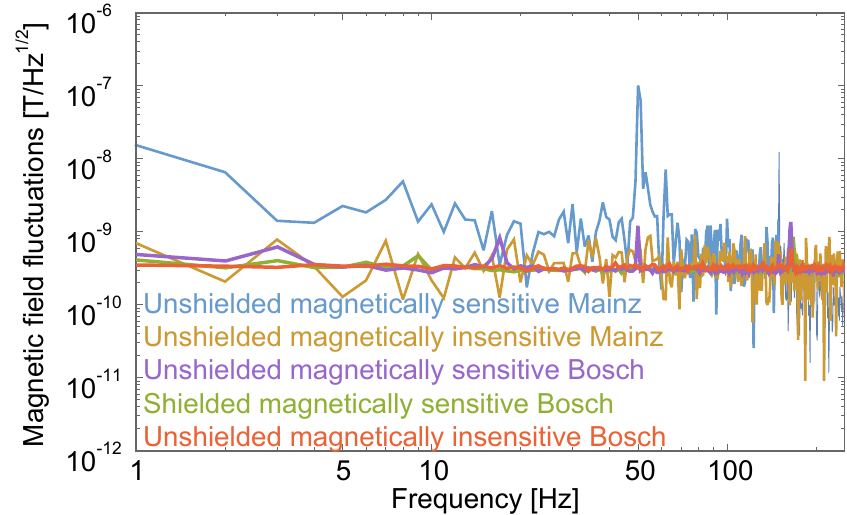}
    \caption{Magnetic field fluctuations measured with the fiberized NV magnetometers. The blue and orange traces depict magnetically sensitive and insensitive data of the Mainz sensor. 
    %while the dashed line notes the photon-shot noise sensitivity limit. 
    The purple and green traces correspond to magnetically sensitive spectra of the Bosch sensor outside and inside a magnetic shield, finally, the red trace corresponds to unshielded magnetically insensitive data.
    The average sensitivity in the 60-90\,Hz band is sub-nT/$\sqrt{\textrm{Hz}}$. The Bosch data are the result of 100\,averages, while the Mainz data is from a single acquisition.}
    \label{fig:3}
\end{figure}

Figure\,\ref{fig:3} shows the magnetic-field-noise spectrum of the sensors. 
%The spectrum for the Mainz sensor is a Fourier transform of the LIA x-output for a measurement time of 1\,s, acquired with data acquisition card, with a reference frequency of f$_{mod}$ \textcolor{blue}{is that true for the Bosch sensor?}.
The blue trace of Fig.\,\ref{fig:3} corresponds to magnetically sensitive data of the Mainz sensor, the orange to magnetically insensitive. %while the dashed line notes the photon-shot noise sensitivity limit calculated for the collected PL, contrast and line-width ($\Delta\nu$) . 
The purple and green trace correspond to magnetically sensitive spectrum of the Bosch sensor outside and inside a magnetic shield, respectively, finally the red trace corresponds to the insensitive plot. The magnetically insensitive spectrum can be obtained if f$_{c}$ is selected to be far from the ODMR features.
The peak at 50\,Hz is attributed to magnetic field from the power line in the lab. The average sensitivity in the 60-90\,Hz area is 453\,pT/$\sqrt{\textrm{Hz}}$ and 344\,pT/$\sqrt{\textrm{Hz}}$ for the Mainz and Bosch sensors, respectively.
%is calculated as 0.5\,nT/$\sqrt{\textrm{Hz}}$. The photon-shot noise sensitivity limit was 11\,pT/$\sqrt{\textrm{Hz}}$. The sensitivity of the sensor was limited by pump-light intensity noise.
The noise traces for the Bosch sensor are based on continuous data series, that are recorded with a sampling rate of 1\,kHz for 100\,s. To calculated the amplitude spectral density (ASD), the data series was split in 100 consecutive intervals, each with a duration of 1\,s. For each interval the ASD was calculated. The depicted is the average of the 100\,ASDs.

\section*{Vector magnetic field sensing}
\subsection*{Data Acquisition}

\begin{figure}
    \centering
    \includegraphics[width=1\columnwidth]{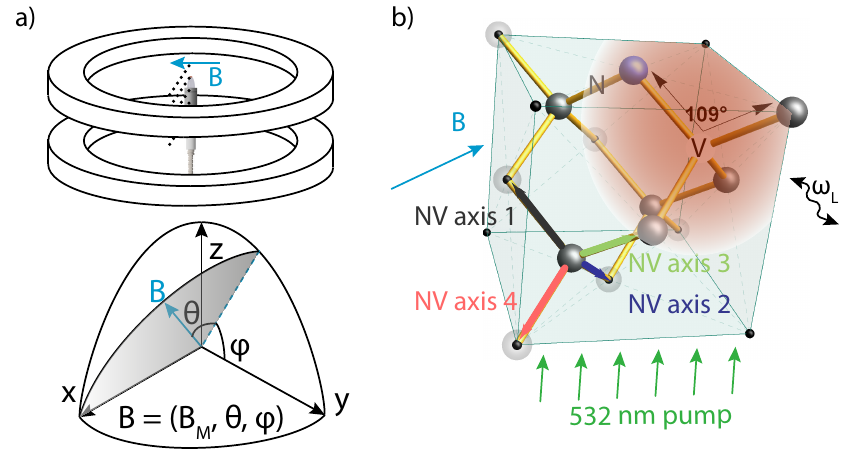}
    \caption{a) The orientation of the diamond sensor in the Halbach array. The orientation of the magnetic field is given in spherical coordinates with angle $\theta$ being the latitude angle with respect to the x-axis and $\phi$ the longitude. The main component of the magnetic field points along the x-axis. b)\,Illustration of the diamond lattice structure indicating the four different NV axes 1,2,3,4 and their relation to the applied magnetic field B. Microwaves (MW) with frequency $\omega_L/\left(2\pi\right)$ are applied to the sample.}
    \label{fig:4}
\end{figure}

As a demonstration of the robustness and portability of our sensors as well as the ability to produce highly resolved magnetic field maps, we select the Mainz sensor to characterize the homogeneity of a custom-made Halbach-array magnet constructed in the Mainz laboratory. The schematic of the magnet is shown in  Fig.\,\ref{fig:4}\,(a) with more details found in Ref.\,\cite{Halbach}. It is a double ring of permanent magnets arranged to generate a homogeneous magnetic field in its inner bore along the radius of the rings. The field outside the construction decays rapidly with distance. We performed ODMR measurements in a $30\times20\,\text{mm}^2$ plane nearly perpendicular to the main magnetic field direction in the center of the Halbach array in steps of 1\,mm and 1.5\,mm in $z$ and $y$-direction, respectively. The experimental procedure to characterize this magnet involves reconstruction of the 3D magnetic field from these ODMR measurements, which we describe in the next part of this paper. The measurements confirmed the homogeneity of the magnetic field of the magnet to be consistent with Hall-probe and NMR measurements, but with a threefold improved field strength resolution, vector information of the magnetic field and sub-mm spatial information of these quantities. The orientation of the fiberized sensor in the Halbach magnet can be seen in Fig.\,\ref{fig:4}\,(a). The magnetic field is given in spherical coordinates with respect to the (100) axis of the diamond $\vec{B}=\left(B_M,\theta,\phi\right)$. The angle between the magnetic field vector and the yz-plane is $\theta$, and $\phi$ is the angle between the projection of $\vec{B}$ in the yz-plane and the y-axis.
The diamond sensor was oriented in the magnetic field such that different resonances were visible.

\begin{figure*}
    \centering
    \includegraphics[width=1\textwidth]{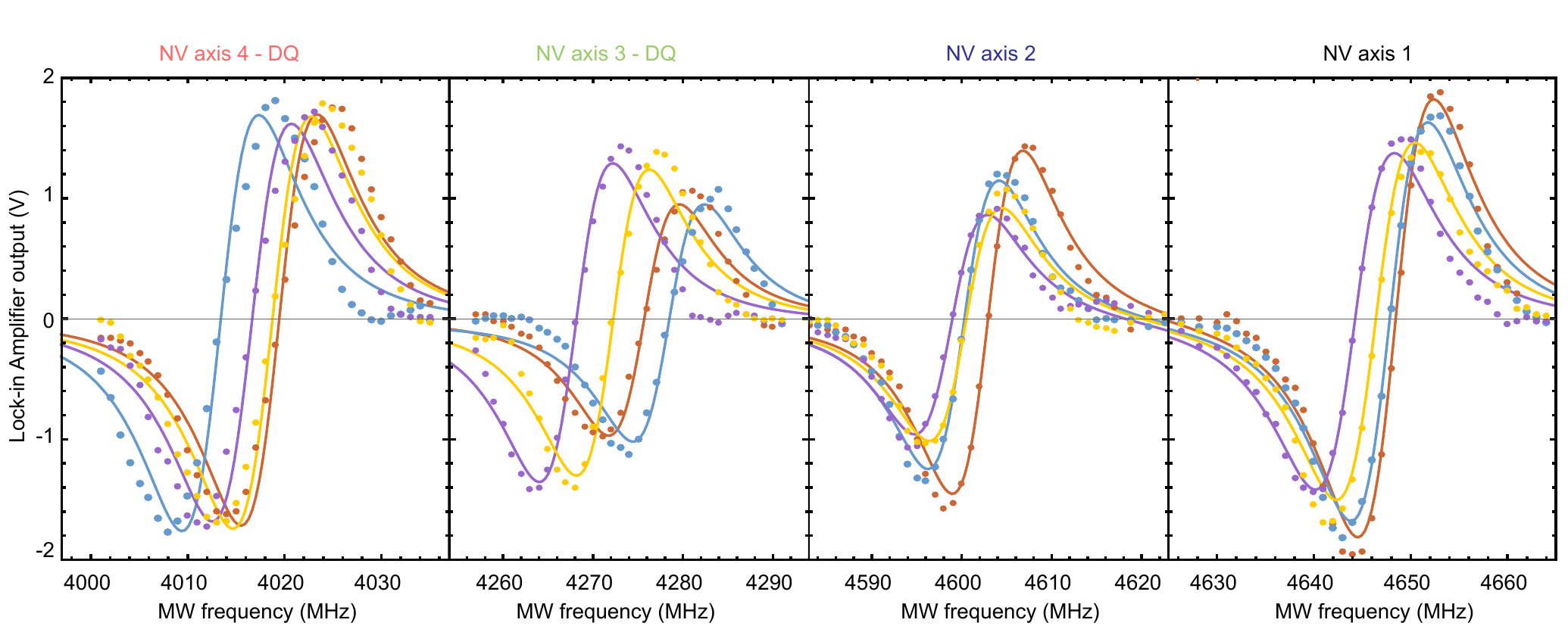}
    \caption{Example data traces of four optically detected resonances at different position near the center of the Halbach magnet. The resonances appear dispersive due to the use of a modulation technique. The resonances are fitted and their center frequencies are extracted. The four different plots correspond to different transitions indicated above the respective plot. The color of the traces indicate the positions at which the data was taken, as seen in Fig.\,\ref{fig:8}.}
    \label{fig:5}
\end{figure*}

For each position in the 30\,$\times$20\,mm$^2$ plane a scan of the applied MW frequency between 3800\,MHz and 5000\,MHz was performed in steps of 1\,MHz. This range was limited by the bandwidth of the MW components. At each frequency we recorded the demodulated PL from the LIA x-output with a data acquisition system and stored each data set with its respective position. To speed up the acquisition and after determining that no features were left out, we just acquired in total 164 frequency values around the expected position of the resonances. In total a frequency scan to determine the four center frequencies lasted 46\,s. This can be dramatically improved by for example applying a frequency lock on the four transitions, respectively.

Figure\,\ref{fig:5} shows an example of collected data in four different points of the scan. The points are noted in Fig.\,\ref{fig:8}\,(b) with different colors. The four different frequency regions correspond to different kind of transitions as noted above the figures. The FWHM linewidth of the observed features is ($11.48\pm0.14$)\,MHz and therefore much larger than the one given above  ($0.35\pm 0.02$)\,MHz for a small background field along one of the NV axis. This is due to the strong transverse field component but also caused intentional via MW power broadening. This simplified the lineshape by suppressing the hyperfine features in the spectrum and therefore the analysis routine. We additionally like to note, that the two first  of the features are double quantum transitions (DQ), i.e. magnetic transitions from the $m_s=-1$ to the $m_s=+1$ state, normally forbidden, but allowed when transverse magnetic field components are present.

Overall the sensor was moved in steps of 1\,mm and 1.5\,mm in $z$ and $y$-direction, respectively, such that the whole area was covered. The resulting frequency maps of the chosen four resonances can be seen in Fig.\,\ref{fig:8}\,(a); they show structures corresponding to the strength of the longitudinal (along the NV axis) and transverse (perpendicular to the NV axis) component of the magnetic field for a given diamond lattice axis. The information contained in these plots is more than sufficient to reconstruct all three vector components of the magnetic field at the position of the sensor.

\begin{figure}
    \centering
    \includegraphics[width=1\columnwidth]{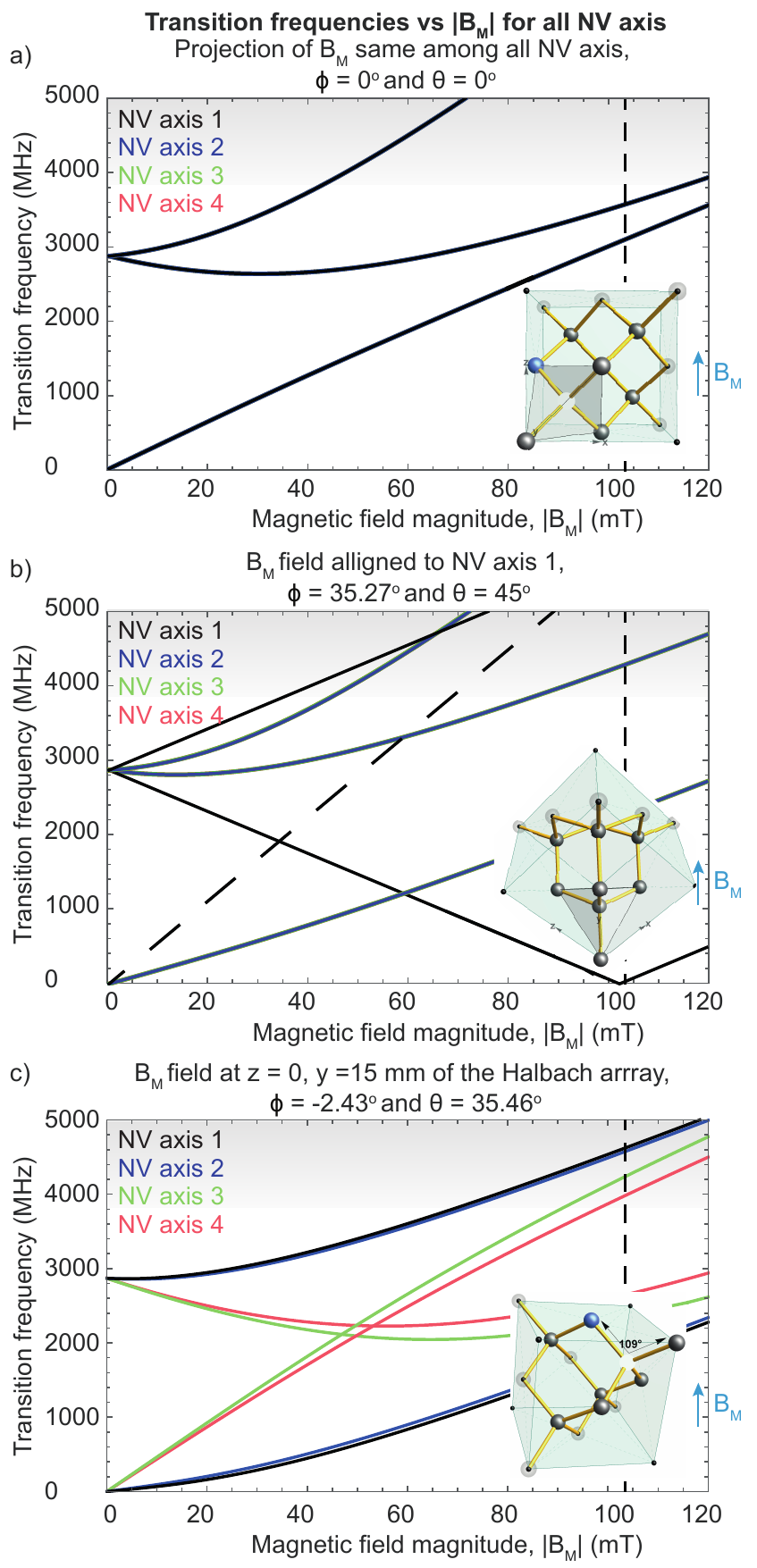}
    \caption{Transition frequencies as a function of $|B_M|$. The traces of the four different NV axes are depicted with colours consistent with those of Fig.\,\ref{fig:4}. The gray area represents our scanning range, the vertical dashed line shows the mean field strength inside the Halbach magnet. a) For a magnetic field perpendicular to the (100) plane: all transitions overlap since the magnetic field projection is the same among all axes. b) For a magnetic field perpendicular to the (111) plane of a diamond: the magnetic field is aligned along NV axis\,1, and the magnetic field projection is the same for the other three axes. The slanted dashed line for NV axis\,1 corresponds to the frequency of the DQ transition which in this case due to the absence of transverse magnetic fields is not allowed. c) For a magnetic field configuration inside the Halbach magnet.}
    \label{fig:versusfield}
\end{figure}

\subsection*{Frequency to vector field conversion}
\begin{figure}
    \centering
    \includegraphics[width=1\columnwidth]{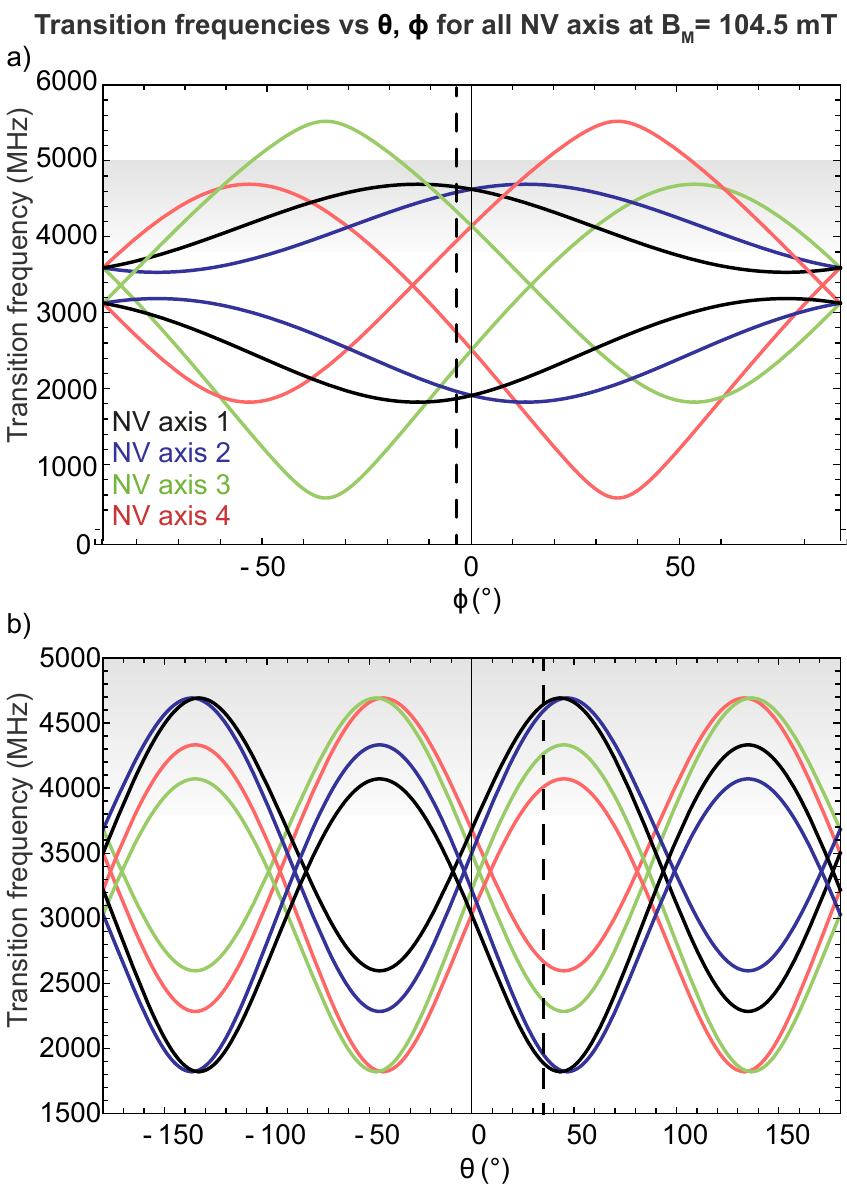}
    \caption{Simulated transition frequencies as a function of angles $\theta$ and $\phi$ for the average magnetic field magnitude $|B_M|=104.5$\,mT inside the Halbach magnet. The gray area represents our frequency scan range, the vertical dashed line indicates the respective average angle of all the measurements. a) Transition frequencies as a function of angle $\theta$ for a fixed angle $\phi=-2.43^{\circ}$. b) Transition frequencies as a function of angle $\phi$ for a fixed angle $\theta=35.46^{\circ}$.}
    \label{fig:versusangle}
\end{figure}

\begin{figure*}
    \centering
    \includegraphics[width=\textwidth]{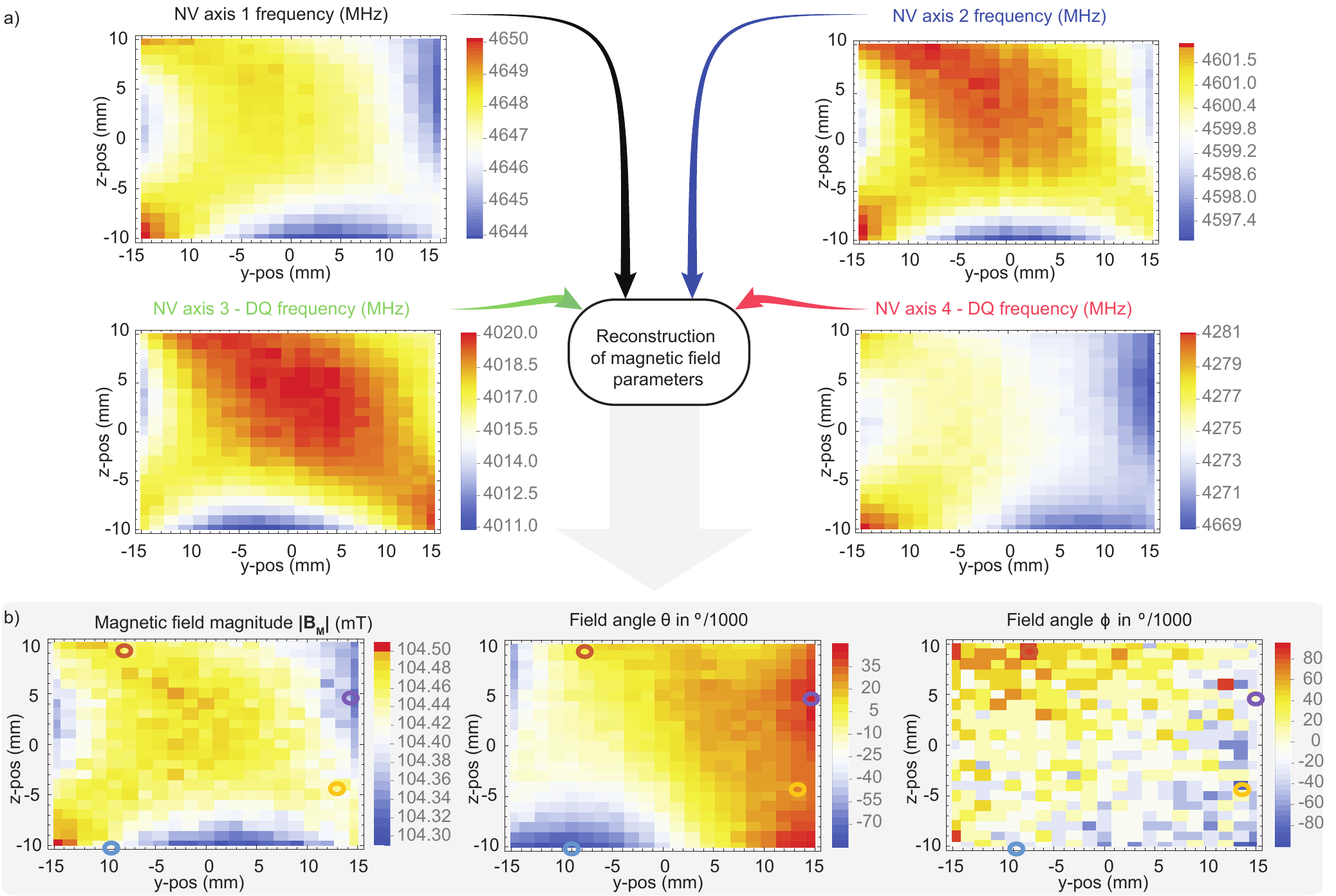}
    \caption[Frequency and field maps]{Figure adapted from Ref.\,\cite{Halbach}. (a) Spatially resolved frequency maps of the individual resonances from Fig.\,\ref{fig:5}\,(a). (b) The data collected are used to reconstruct the vector magnetic field in spherical coordinates as defined in Fig.\,\ref{fig:4}\,(a), the four colored circles note the positions at which data for Fig.\,\ref{fig:5} are taken.}
    \label{fig:8}
\end{figure*}

After acquiring MW frequency scans for different positions, the data were fitted with the sum of four derivatives of Lorentzians. The four center frequencies, four amplitudes and a combined linewidth were fit parameters, an example  of the data and fits can be seen in Fig.\,\ref{fig:5}. By matching these frequencies to the positions of the 3D translation stages we can make frequency maps as shown in Fig.\,\ref{fig:8}\,(a).

To proceed with the analysis and construction of the magnetic field maps we note that the four different features presented in Fig.\,\ref{fig:5} originate from the different NV orientations in the diamond crystal. The positions of these features are related to the alignment of the magnetic field to the NV axis, as well as its amplitude. If the magnetic field is along the NV axis (longitudinal) we observe the $m_s=0\rightarrow\,m_s=\pm1$ transitions; if there is an additional magnetic field component orthogonal to the NV axis (transverse), transitions between the $m_s=-1$ and $m_s=+1$ are allowed. We call these double quantum (DQ) transitions. For a given magnetic field direction at the position of the sensor, both, the longitudinal and transverse components of the magnetic field are present. %The strength of the magnetic field multiplied by the gyromagnetic ratio would result then in a frequency shift. 
Due to mixing caused by the transverse magnetic field component, the NV gyromagnetic ratio depends on the background field strength. 

As a result, to match the above frequencies to magnetic fields we create four vectors describing the NV axes. We use parameters $\beta$ to describe the angle between NV axis and magnetic field and $\Omega$ for the field strength. With these parameters we derive a formula which describes the transition frequencies of the spin states, from m$_s$\,=\,0 to m$_s$\,=\,-1, as well as m$_s$\,=\,-1 to m$_s$\,=\,+1\,\cite{Simon}. Here, we neglect strain and electric field effects.  

 The frequency of the resonances depends on the magnitube of the magnetic field B$_M$, as well as the angles $\theta$ and $\phi$ of the diamond orientation with respect to the magnetic field. Figure\,\ref{fig:versusfield} shows the resonance frequency dependence as a function of magnetic field magnitude for different fixed angles $\theta$ and $\phi$. Figures\,\ref{fig:versusfield}\,(a,b) represent a magnetic field perpendicular to common diamond surface cuts, (100) plane for (a) and (111) plane for (b), (c) depicts a configuration inside our Halbach magnet. Figure\,\ref{fig:versusangle} shows the dependence of the transition frequencies on the angles $\theta$ and $\phi$ for a magnetic field matching the mean magnitude inside the Halbach magnet. In Fig.\,\ref{fig:versusangle}\,(a) $\phi$\,=\,-2.43$^{\circ}$ and $\theta$ is varied. In Fig.\ref{fig:versusangle}\,(b) $\theta$\,=\,35.46$^{\circ}$ and $\phi$ is varied.

In both Fig.\,\ref{fig:versusfield} and Fig.\,\ref{fig:versusangle} the four different NV axes are depicted with colours matching those in Fig.\,\ref{fig:4}. We observe that in Fig.\,\ref{fig:versusfield}\,(a) all the transitions overlap, as the projection of the magnetic field is the same among all axes. In Fig.\,\ref{fig:versusfield}\,(b) the magnetic field is aligned along the NV axis 1, and its projection is the same for the other three axes, which appear overlapped in the figure. The dashed line for NV axis 1 corresponds to the frequency of the DQ transition which in this case is not allowed, as there is no transverse field on NV axis 1.
Figure\,\ref{fig:versusfield}\,(c) represents a case we encounter while measuring the field inside our Halbach magnet. In Fig.\,\ref{fig:versusangle}, we note the symmetry between NV axes 1,2 and 3,4 which arises because of the crystallographic structure of the diamond. We can use the information of the above plots to translate the frequency into magnetic field. 
%Figure\,\ref{fig:5}\,(b) shows an example plot of all allowed transition frequencies for all axes in a diamond as a function of the magnetic field magnitude. The, angle $\beta$ for this plot is fixed. 
%We can see the different NV axes display different gyromagnetic ratios, which can be found from the slope of the curves. 
%The dashed lines in Fig.\,\ref{fig:5}\,(b) correspond to measured microwave frequencies inside the Halbach magnet.

After identifying the possible transition frequencies, we use a numerical method to translate the frequencies into magnetic field direction and strength. For this method we restrict the magnetic field strength values to within 10\% of what the estimated magnetic field produced by our Halbach array is. Once the magnetic field is calculated for all the different frequencies we express it in polar coordinates.
%One can argue for the uniqueness of the field via the C$_3v$ symmetry to the segment of a sphere. This renders the system invariant and the eigenvalues in each of those segments, correlated to each other by a phase.
When the initial values for direction and magnitude of the field are approximately known, we find that the reconstruction is unique.
We choose $B_M$ to describe the strength of the field, angle $\theta$ the latitude angle with respect to the x-axis and $\phi$ the longitude angle as shown in Fig.\,\ref{fig:4}\,(a). Finally, we make a list of these parameters and plot them as shown in Fig.\,\ref{fig:8}\,(b). In Fig.\,\ref{fig:8}\,(b) we included four different coloured circles to represent the positions at which data for Fig.\,\ref{fig:5} are taken.  The measured averages of angles $\theta$ and $\phi$ in the central $10\times5$~mm$^2$ (y-x) of the Halbach magnet are $\left(-2.433\pm0.009\right)^\circ$ and $\left(35.764\pm0.018\right)^\circ$, respectively, which is consistent with the orientation of the diamond lattice (cf. Fig.\,\ref{fig:4}) relative to the coordinate system of the magnet. 
%\begin{align*}
%\beta _1 &= \beta\\
%\beta _2&=\cos ^{-1}\left(\frac{1}{3} \left(-\sqrt{2} \sin (\beta \left(\sqrt{3} \sin (\alpha)+\cos (\alpha)\right)-\cos (\beta)\right)\right)\\
%\beta _3&=\cos ^{-1}\left(\frac{1}{3} \left(\sqrt{6} \sin (\alpha ) \sin (\beta )-\sqrt{2} \cos (\alpha ) \sin (\beta )-\cos (\beta )\right)\right)\\
%\beta _4&=\cos ^{-1}\left(\frac{1}{3} \left(-\sqrt{2} \sin (\beta ) \left(\sqrt{3} \sin (\alpha )+\cos (\alpha )\right)-\cos (\beta )\right)\right)\\
%\end{align*}

\section*{Conclusion and Outlook}
We constructed two fiberized NV based magnetic field sensors. The sensors feature sub-nT/$\sqrt{\textrm{Hz}}$ magnetic field sensitivity with high 
%a 0.5\% 
PL-to-pump-light ratio. The components used to make the sensors are commercially available which makes their construction easy and reproducible. The design for the Mainz sensor was made in such a way as to allow free access to the front side of the diamond as well as for robustness, portability, and small size. 

With one of the main challenges for NV magnetometry, especially fiberized, being the low photon-collection efficiency, which leads to poor photon-shot-noise limited sensitivity, the 0.5\% PL-to-pump-light ratio collection efficiency, demonstrated in the Mainz sensor, can be used to achieve even higher sensitivity, if the effect of other noise sources is minimized. This ratio could also be further optimized by implementing side collection on the diamond\,\cite{Sage2012,2019BarryNVReview}, as well as highly reflective coatings on the opposite (front) side of the diamond. These features were not implemented in this sensor in order to keep all of its components commercial and easily accessible. 

The access to the front side of the diamond allows for close proximity to magnetic field sources, which in turn leads to high spatial resolution. The proximity is currently limited to about 300\,$\mu$m by MW wire on top of the diamond. It could be optimized further by using a thinner diamond sample, a sample with shallow implanted NV centers or a thinner MW wire such as, for example, a capton-tape printed circuit board, all depending on the intended application, as these changes influence the sensitivity of the sensor. Such optimization would be especially beneficial for measurements of dipole fields or measurements that require high spatial resolution. 

The robustness and portability of the sensors also make them attractive for mapping larger areas, or on-the-move measurements of magnetic fields. These features combined with the small size of the sensors suggest a use for measuring in hard-to-access places or small areas where maneuverability of the sensor would be required, e.g. endoscopic measurements inside the human body. 

As a demonstration of the Mainz sensor we measured the magnetic field of a custom-designed Halbach array. For these measurements, the small fiberized diamond sensor showed a mm-scale spatial resolution, angular resolution of the magnetic field vector extracted in the measurements were $9\times10^{-3}$ degrees and $18\times10^{-3}$ degrees for angles $\theta$ and $\phi$, respectively. We note here, that the corresponding displacement due to such a rotation of the diamond sensor (considering a $(0.5\times0.5)$~mm$^2$ square cross section) could optically not be resolved.
%with a volume of $500^3\,\mu\text{m}^3$ 

\section*{Acknowledgment}
This work is supported by the EU FET-OPEN Flagship Project
ASTERIQS (action 820394), the German Federal Ministry of Education and Research (BMBF) within the Quantumtechnologien program (grants FKZ 13N14439 and FKZ 13N15064), the Cluster of Excellence “Precision Physics, Fundamental Interactions, and Structure of Matter” (PRISMA+ EXC 2118/1) funded by the German Research Foundation (DFG) within the German Excellence Strategy (Project ID 39083149). We acknowledge Fedor Jelezko for fruitful discussions.

\bibliographystyle{ieeetr}
\bibliography{./literature/Literature.bib}

\end{document}